\documentstyle[alleqno,epsfig]{mn}
\newif\ifAMStwofonts
\AMStwofontstrue
\ifoldfss

  \ifCUPmtlplainloaded \else
    \NewTextAlphabet{textbfit} {cmbxti10} {}
    \NewTextAlphabet{textbfss} {cmssbx10} {}
    \NewMathAlphabet{mathbfit} {cmbxti10} {}
    \NewMathAlphabet{mathbfss} {cmssbx10} {}
  \fi
  \ifAMStwofonts
    \ifCUPmtlplainloaded \else
      \NewSymbolFont{upmath} {eurm10}
      \NewSymbolFont{AMSa} {msam10}
      \NewMathSymbol{\upi}     {0}{upmath}{19}
      \NewMathSymbol{\umu}     {0}{upmath}{16}
      \NewMathSymbol{\upartial}{0}{upmath}{40}
      \NewMathSymbol{\leqslant}{3}{AMSa}{36}
      \NewMathSymbol{\geqslant}{3}{AMSa}{3E}

      \let\leq=\leqslant 
       
    \fi
  \fi
\fi
\ifnfssone
   \newmathalphabet{\mathit}
  \addtoversion{normal}{\mathit}{cmr}{m}{it}
  \addtoversion{bold}{\mathit}{cmr}{bx}{it}

   \newmathalphabet{\mathbfit} 
  \addtoversion{normal}{\mathbfit}{cmr}{bx}{it}
  \addtoversion{bold}{\mathbfit}{cmr}{bx}{it}
  \newmathalphabet{\mathbfss} 
  \addtoversion{normal}{\mathbfss}{cmss}{bx}{n}
  \addtoversion{bold}{\mathbfss}{cmss}{bx}{n}
  \ifAMStwofonts
    \ifCUPmtlplainloaded \else
      \UseAMStwoboldmath
      \makeatletter
      \new@mathgroup\upmath@group
      \define@mathgroup\mv@normal\upmath@group{eur}{m}{n}
      \define@mathgroup\mv@bold\upmath@group{eur}{b}{n}
      \edef\UPM{\hexnumber\upmath@group}
      \new@mathgroup\amsa@group
      \define@mathgroup\mv@normal\amsa@group{msa}{m}{n}
      \define@mathgroup\mv@bold\amsa@group{msa}{m}{n}
      \edef\AMSa{\hexnumber\amsa@group}
      \makeatother
      \mathchardef\upi="0\UPM19
      \mathchardef\umu="0\UPM16
      \mathchardef\upartial="0\UPM40
      \mathchardef\leqslant="3\AMSa36
      \mathchardef\geqslant="3\AMSa3E

      \let\leq=\leqslant 
       
    \fi
  \fi
\fi
\ifnfsstwo

  \DeclareMathAlphabet{\mathbfit}{OT1}{cmr}{bx}{it}
  \SetMathAlphabet\mathbfit{bold}{OT1}{cmr}{bx}{it}
  \DeclareMathAlphabet{\mathbfss}{OT1}{cmss}{bx}{n}
  \SetMathAlphabet\mathbfss{bold}{OT1}{cmss}{bx}{n}
  \ifAMStwofonts
    \ifCUPmtlplainloaded \else
      \DeclareSymbolFont{UPM}{U}{eur}{m}{n}
      \SetSymbolFont{UPM}{bold}{U}{eur}{b}{n}
      \DeclareSymbolFont{AMSa}{U}{msa}{m}{n}
      \DeclareMathSymbol{\upi}{0}{UPM}{"19}
      \DeclareMathSymbol{\umu}{0}{UPM}{"16}
      \DeclareMathSymbol{\upartial}{0}{UPM}{"40}
      \DeclareMathSymbol{\leqslant}{3}{AMSa}{"36}
      \DeclareMathSymbol{\geqslant}{3}{AMSa}{"3E}

      \let\leq=\leqslant 
       
    \fi
  \fi
\fi 
\ifCUPmtlplainloaded \else
  \ifAMStwofonts \else
    \def\upi{\pi}
    \def\umu{\mu}
    \def\upartial{\partial}
  \fi
\fi
\title{Exact Optics: A unification of optical telescope design.}
\author[D. Lynden-Bell]
       {D. Lynden-Bell \\
        Institute of Astronomy, Cambridge, U.K.}
\date{Accepted 
      Received ;
      in original form }
\pagerange{\pageref{firstpage}--\pageref{lastpage}}
\pubyear{2001}

\begin{document}
\maketitle
\label{firstpage}
\begin{abstract}
A perfect focus telescope is one in which all rays parallel to the
axis meet at a point and give equal magnification there.  It is shown
that these two conditions define the shapes of both primary and
secondary mirrors.  Apart from scale, the solution depends upon two
parameters, $s$, which gives the mirror separation in terms of the
effective focal length, and $K$, which gives the relative position of
the final focus in that unit.  The two conditions ensure that the
optical systems have neither spherical aberration nor coma, no matter
how fast the $f$ ratio.  All known coma--free systems emerge as
approximate special cases.  In his classical paper, K. Schwarzschild
studied all two mirror systems whose profiles were conic sections.  We
make no such a priori shape conditions but demand a perfect focus
and solve for the mirrors' shapes.  
\end{abstract}
\section{Introduction}
When the author saw the optical design for the corrector of the
Hobby--Eberly Telescope, which involved four reflections after the
primary and produced a four minute of arc field, he decided that it
ought to be possible to do better.  This led him to study optics from
first principles.  The methods used have much in common with those of
Descartes (1634), (see Smith 1925) but the author has one signal
advantage in that Newton's calculus is known now, whereas Descartes
(1596--1650) had to rely on geometry alone; (indeed it is of some
interest that Descartes seems to have invented analytical geometry in
order to solve optical problems).  Optics and optical phenomena have
stimulated work by many great scientists besides Descartes.  Galileo
(1564--1642), Mersenne (1588--1650)(1636), Fermat (1601--1665),
Huygens (1629--1695), Hooke (1635--1702), Newton (1642--1727)(1704),
W. Herschel (1738--1822), Young (1773--1829)(1802), Fresnel
(1788--1827), Hamilton (1788--1856)(1824, 1830, 1832, 1931), Foucault
(1819--1868), Zeiss (1816--1888), Seidel (1821--1896)(1856), Maxwell
(1831--1879), Abbe (1840--1905), Michelson (1852--1931), Schwarzschild
(1873--1916)(1905), Schmidt (1879--1935)(1931), Einstein (1879--1955),
Zernicke (1888--1966) made major advances in our understanding.  The
work of Lord Rosse (1800--1867), Ritchey (1864-1945) and Chr\'{e}tien
(1879--1956)(1922) much improved the optics of large telescopes.

In past centuries aspheric surfaces were very
expensive and difficult to make, so attention was concentrated on
spherical surfaces or those whose profiles were conic sections (see
e.g., Schwarzschild (1905)).  
The coming of computers has enabled the designer to evaluate the
performance of even very complicated optics with ease.  Thus computers
may be programmed to optimise a design according to whatever criteria
are chosen.  This has revolutionised optical design.  In reality the
design is computed by directed trial and error.
In the hands of an experienced optical
designer this is a very powerful and adaptable method which will
doubtless remain the prime tool for the foreseeable future, see e.g.,
Willstrop, R.V. (1987) who gave a fine wide--field telescope design.
Angel's wide--field survey telescope is based on a development of this
design with smaller field and a smaller hole in the primary.  I
understand that this is now to be built with an eight metre primary mirror.

Traditional optical theory expands all the trigonometric functions in
powers of the angles.  Baker (1940) and Burch (1942)
implemented Petzval's theory 
to eliminate the lowest order astigmatism, and the book by
Korsch (1992) gives an account of further developments in that
direction.  Computer based ray--tracing developed by Wynne (1959, 1974) led
to fine designs for multi-element glass prime focus correctors which
gave larger fields to a whole generation of optical telescopes.
However, the design of aerials for centimeter and millimeter radio
astronomy and communications led to a requirement for very fast
designs where the angles were not small.  The coming of computers has
allowed such designs to be investigated numerically and this led  to a
new flowering of computationally based optics.  The book by Mertz
(1996) gives original, fast and imaginative optical designs.  Methods
for fast optical designing are given in the book by Cornbleet (1994).
This is no place to discuss the myriad of designs produced by computer
optics because the paper is devoted to analytical mathematical
theory.

Quite recently new optical fabrication techniques have made it
possible to produce mirrors of any desired profile, although those
with axial symmetry are still much cheaper.  This makes the study of
systems with mirrors which may be of any shape especially
topical. Whereas the earliest workers prescribed the shapes and then
asked what the system would do, cataloging its failures as
``aberrations'', the modern ray--tracing methods are really solving an
inverse \linebreak problem in which the mirror shapes are varied until the
desired performance is achieved as closely as possible.  Here we shall
use analytical methods but leave the shapes of the two mirrors to be
determined so as to give an exact on--axis focus near which all rays
give equal magnification.  We show that this problem may be solved
exactly yielding a two parameter family of exact ray--optical solutions
even for very fast $F$ ratios.  Some of our solutions are more
appropriate for spectrograph cameras than for telescopes. Others, with
exact foci at which the rays enter over a hemisphere or more, may be
appropriate for solar furnaces or lighthouses.  In discovering exact
formulae for all two mirror telescopes/cameras with neither spherical
aberration nor coma the difficult part is not in proving the theorems
but in finding the right variables so that the problem can be solved
analytically in parametric form.  Directed trial and error by a
persistent mathematician who feels the problem is simple enough to
have a nice solution, here replaces the directed trial and error of
the optical designer with a computer.  However, although the
analytical method covers all cases at once, it is confined to perfect
images at and close to the optic axis, whereas the main interest lies
in the breadth of field off the axis.  The author has gained much from
the tutelage of an expert on optics, Dr R.V. Willstrop, and he has
used his computer ray--tracing methods to show that these designs give
as good off--axis performance as those designed with computers.

Although there are some new designs to be found among our two
parameter set of solutions, most of the useful designs were discovered
long ago.  In this respect, the present paper may be regarded as a
unification of all those optical designs into analytical formulae.
These formulae have the small advantage of giving `perfect' on axis
performance.  The reason we have concentrated on the removal of
spherical aberration and coma is that these aberrations depend on the
lowest powers of the angle off axis, $\alpha$.  They are therefore the
dominant aberrations close to the axis.  Table 1 
gives the behaviour of the aberrations with that angle and with the
$F$ ratio, $F = $ focal length/aperture.  

\begin{table}
\caption{Dependence of transverse angular aberrations on field angle
  $\alpha$ and $F$-ratio.}
\begin{tabular}{ll}
spherical aberration    & $\alpha^0 F^{-3}$\\
coma                    & $\alpha^1 F^{-2}$\\
astigmatism             & $\alpha^2 F^{-1}$\\
field curvature         & $\alpha^2 F^{-1}$\\
distortion              & $\alpha^3 F^0$\\
& \\
\end{tabular}
\hrule
\end{table}

Whereas our systems are free of the first two aberrations by design,
some of them are also free of astigmatism.  It may be argued that
digital recorders may in principle be made to fit field curvature, and
distortion is readily removed in the computer.  

Even on axis, the rays emerging from some achromatic axially symmetrical
optical train do not generally give a good focus; rather those rays at
an angle $2\theta$ to the axis will cross it at some point $x(\theta)$
which depends on $\theta$.  The function $x(\theta)$ plays a prominent
role in our analysis so we call it the defocusing function.   Only
when it is constant do all the rays pass through a focus without any
corrector. Such systems are said to have no spherical aberration.

In section 2 we give a Lemma that shows how to correct a given set of
rays for spherical aberration.  This Lemma can be useful on its own
when only a point focus is needed, so coma is unimportant.

We show that rays with a known defocusing function may be
redirected to an exact focus at any chosen point, $x_f$, on axis by a
corrector mirror whose pole is at $x_c$.  This mirror's shape and size
are given explicitly in terms of the defocusing function $x(\theta)$
and the parameters $x_f$ and $x_c$.  We also show that the corrector
mirror's shape is completely determined by the uniting function
$U(\theta)$ which unites the dependency on $x(\theta)$ with that on
$x_f$ and $x_c$.

In section 3 we show that any given defocusing function can arise
from a single primary mirror whose pole is at any chosen $x_p$.  Thus
$x(\theta)$ and $x_p$ together totally determine the shape, size and
position of the primary mirror.  We go on to show that any two mirror
telescope is characterised by specifying just the uniting function
$U(\theta)$ and the mirror separation $x_p - x_c$.  However, most such
telescopes will be almost useless save for limited point spectroscopy
because the rays that come to the final focus originating from
different rings of the primary mirror will arrive carrying different
magnifications.  Objects a little off axis will therefore suffer from
coma which can be severe.

Many years ago, Abbe (1873) (see Jenkins \& White 1957) gave the
criterion for eliminating coma near the axis.  The rays arriving at
focus when projected back must meet the corresponding incoming
parallel rays on a sphere centred at the focus.

In section 4 we show that Abbe's condition used in conjunction with
the results of sections 2 and 3 gives a differential equation which we
solve to determine the shapes of both primary mirror and corrector.
Thus we have determined all perfect--focus two mirror telescopes.  The
resulting systems depend on three parameters: 1. The scale; 2. A
dimensionless ratio $s$ that determines the separation of the two
mirrors in units of effective focal length of the whole system; 3. The
dimensionless ratio $K$ that measures the distance of final focus from
the secondary measured in that unit. Thus section 4 proves the main
theorem, and gives the shapes of the mirror systems parameterised by
$s$ and $K$.

This paper derives and explores the properties of such systems {\bf on
axis}.  Their properties off axis are best explored by ray tracing.
The accompanying paper describes the results of such studies.

 We thank the referee for pointing out, that the problem of
designing two aspheric surfaces has been considered more concisely
by Born \& Wolf (1999) and is the subject of their paragraphs 4.10.1 and
4.10.2. Indeed they treat both the lens and the mirror cases, but
paragraph 4.10.2 ends somewhat lamely by giving two complicated
simultaneous first order non-linear differential equations with their
boundary conditions and saying  that they may be integrated numerically
case by case. The advance made here is that for two mirrors new variables
are found in terms of which the differential equations are integrated
analytically to give the general solution for the mirrors' shapes.
Exploration of the full panoply of solutions, for all $s$ all $K$ and all
$F$-numbers, is therefore made easy. The actual mirror shapes found for very
fast systems have cusps and asymptotes which might prove awkward to
compute numerically  over the full range in the variables of Born \& Wolf.
However fore-warned is fore-armed and given the answers here derived those
who prefer totally numerical work will no doubt be able now to reproduce
each particular case in turn, to the accuracy of the computation.

\section{A spherical aberration corrector for any rays with an axis}
Let the rays emerging from some optical train at an angle $2\theta$ to
the axis intersect it at some point $x(\theta)$ as in Figure 1.  We
wish to find the shape of a corrector mirror (with pole at $x_c$)
which will bring rays of all $\theta$ to an exact focus at some prescribed
point on axis, $x_f$.
\subsection*{Lemma}
Spherical aberration can be eliminated by a corrector mirror whose
shape $R(X)$ is given parametrically as $R(\theta),\ X(\theta)$, where
\begin{eqnarray*}
R(\theta) =U\arcmin(\theta)\left [\sin 2\theta \left ({1-\tau^2
\over 2\tau}\right ) - \cos 2\theta \right ]\ ,
\end{eqnarray*}
\begin{eqnarray*}
X(\theta) =x_f -U\arcmin(\theta)\left [\cos 2\theta \left ({1-\tau^2
\over 2\tau}\right )+ \sin 2\theta \right ]\ ,
\end{eqnarray*}
where the uniting function $U(\theta)$ is given by
\begin{eqnarray*}
U(\theta)=\int^\theta_0\left [x_f -x(\theta)\right] \sin 2 \theta
d\theta - (x_f - x_c) \ 
\end{eqnarray*}
\begin{eqnarray*}
U\arcmin(\theta)=dU/d\theta = \left [x_f -x(\theta)\right ]\sin
2\theta\ ,
\end{eqnarray*} 
and
\begin{eqnarray*}
\tau = -{\scriptscriptstyle{1\over 2}}U\arcmin(\theta)/U(\theta)\ .
\end{eqnarray*}
Notice that the shape of the mirror is determined completely once the
uniting function $U(\theta)$ is known but for the position of the mirror
we need to know $x_f$ as well as the function $U(\theta)$
\begin{figure}
\epsfig{figure=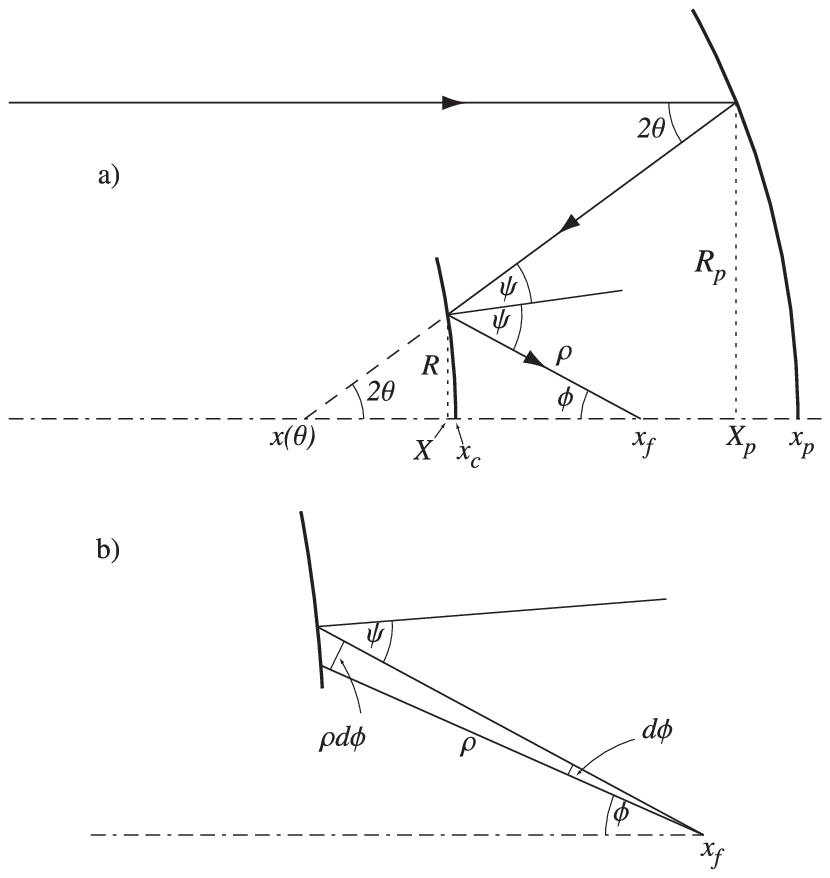}
\caption{\eject
{\bf a)} The simplest layout of a two mirror system giving the
notation. \eject 
{\bf b)} Detail of the corrector mirror with the angles involved.}
\label{fig1ab.eps}
\hrule
\end{figure}
\subsection*{Proof}
From the geometry of figure 1a, a point $(X,R)$ on the corrector
mirror must obey
\[
R/[X-x(\theta)] = \tan 2\theta, \label{eq1}
\]
\[
x_f -X          = \rho \cos \phi, \label{eq2}
\]
\[
R               =  \rho \sin \phi,      \label{eq3}
\]
Eliminating $R$ and $X$ from (\ref{eq1}), (\ref{eq2}) and (\ref{eq3})
we find, 
\[
\rho(\sin \phi \cos 2\theta + \cos \phi \sin 2\theta) = \left [x_f
-x(\theta)\right ] \sin 2 \theta \equiv g(\theta)\label{eq4}\ .
\]
As $x(\theta)$ and $x_f$ are both known, $g(\theta)$ may be thought of
as given.  Writing c.f. fig.1a 
\[
2\theta + \phi = 2 \psi, \label{eq5}
\]
(\ref{eq4}) simplifies to give,
\[
\rho \sin 2 \psi = g(\theta). \label{eq6}
\]
The gradient of the corrector mirror can be described in $\rho,\phi$
polar coordinates at the focus as in figure 1b.
\[
\rho^{-1}d\rho/d\phi = \tan \psi = \tau \ . \label{eq7}
\]
Differentiating (\ref{eq5}) with respect to ${\rm ln}\ \rho$, 
\begin{eqnarray*}
2d\theta/d{\rm ln}\rho (1-d\psi/d\theta)= - \rho d\phi/d\rho = -
1/\tau\ , 
\end{eqnarray*}
so
\[
d{\rm ln}\rho/d\theta = - 2\tau (1-d\psi/d\theta)\ . \label{eq8}
\]
But, by taking logs and differentiating (\ref{eq6})
\begin{eqnarray*}
d{\rm ln}\rho/d\theta +\left [\left (1-\tau^2 \right )/\tau \right ]
d\psi/d\theta = g\arcmin(\theta)/g\ .
\end{eqnarray*}
Substituting for $d{\rm ln}\rho/d\theta$ from (\ref{eq8}) and multiplying by
 $g/\tau$
\begin{eqnarray*}
g\arcmin/\tau - \left [\left (1+\tau^2\right )/\tau^2\right ]\left
(d\psi/d\theta\right )g = -2g;
\end{eqnarray*} 
But from (\ref{eq7}), $(1+\tau^2)d\psi/d\theta$ is $d\tau/d\theta$ so
the left hand side is just $d (g/\tau)/d\theta$.  We integrate and
obtain
\[
-{\scriptscriptstyle{{1\over 2}}} (g/\tau) = U(\theta)\label{eq9}\ ,
\] 
where $U = \int^\theta_0 g(\theta)d\theta -C$ and $U\arcmin(\theta)=g$. 

$C$ is the constant of integration which we determine by taking $\psi$
small then $\rho\rightarrow \frac{1}{2}g/\tau = - U\rightarrow C$.
Furthermore from figure 1, $\theta$ is then small and $\rho \rightarrow
x_f - x_x$.  Hence $C=x_f-x_c$ so $U$ is totally determined by
$x(\theta), x_f\ {\rm and }\ x_c$, 
\[
U=\int^\theta_0 [x_f - x(\theta)]\sin 2\theta d\theta -(x_f -x_c)\ .
\label{eq10}
\]
Since $U(\theta)$ is totally known in terms of given quantities, it is
useful to express the final results in terms of $U$.  From (\ref{eq9})
\[
\tau = - {\scriptscriptstyle{\frac{1}{2}}} U\arcmin/U \ , \label{eq11}
\]
so $\tau \equiv \tan \psi$ is known, so from (\ref{eq6}),
\[
\rho = U\arcmin/\sin 2 \psi =
{\scriptscriptstyle{\frac{1}{2}}}(1+\tau^2)U\arcmin/\tau = -
(U^2+{\scriptscriptstyle{\frac{1}{4}}}U^{\prime 2})/U\ . \label{eq12}
\]
If we put $T = \tan(\phi/2)$ and $t=\tan\theta$ then 
\[
T=\tan (\theta - \psi) = (t-\tau)/(1+t\tau)\ , \label{eq13}
\]
so $T$ is a known function of $\theta$ since (\ref{eq11}) gives
$\tau(\theta)$.  The equations that determine the corrector's shape are
now parametric equations in terms of $\theta$ with $U(\theta)$ via
(\ref{eq10}), $\tau$ via (\ref{eq11}), $\rho$ via (\ref{eq12}) and $T$ via
(\ref{eq13}),
\[
R = \rho\ 2T/(1+T^2)\ , \label{eq14}
\] 
and
\[
X = x_f - \rho (1-T^2)/(1+T^2)\ , \label{eq14a}
\]
or equivalently, 
\[
R(\theta) =U\arcmin(\theta)\left [\sin 2\theta \left ({1-\tau^2
\over 2\tau}\right ) - \cos 2\theta \right ], \label{copyeq1}
\]
\[
X(\theta) =x_f -U\arcmin(\theta)\left [\cos 2\theta \left ({1-\tau^2
\over 2\tau}\right )+ \sin 2\theta \right ]\label{copyeq2}\ .
\]
Q.E.D.

Notice that these equations give $R(\theta)$ and $X(\theta)$ uniquely
once $x(\theta)$ and the parameters $x_f$ \& $x_c$ are
known. Equivalently, if $U(\theta)$ is taken as known, then the shape of
the corrector mirror is determined by (\ref{eq11}), (\ref{copyeq1}),
(\ref{copyeq2}) and different values of
$x_f$ give the same mirror displaced provided that the same function
$U(\theta)$ is used.  In this respect $U(\theta)$ is more closely
related to the corrector's shape than is the defocusing function
$x(\theta)$.  Notice also that systems for which the corrector has the
same intrinsic shape but are scaled up by a factor $\lambda$ have
$\lambda U$ replacing $U$.  So in this respect too the function $U$
characterises the corrector mirror.  It is therefore sensible to use
$U(\theta)$ rather than $x(\theta)$ where this is possible.

In any particular case we plot $R(\theta)$ against $X(\theta)$ to get
the corrector's shape.

Post--focus correctors are already included in the analysis if we allow for
$\phi$ being of the opposite sign to $\theta$ and $R<0$.  Since the
solution -- (\ref{eq14}) (\ref{eq14a}) -- is given parametrically, we may
expect to encounter the cusps and folds of catastrophe theory.
Indeed, some of the strangest mirrors come from these.

This Lemma is formally stated here because it is needed in the proof
of the main theorem in section 4.  The Lemma is not new.  Cornbleet
(1994) states that $x(\theta)$ determines the mirror's shape, and the
principles on which the Lemma is based are clearly stated by Descartes
(1634). 
\subsection*{Example -- Correctors for a Spherical Primary Mirror} 
Parallel rays fall on a concave spherical mirror of radius of
curvature $a$.  The reflected rays at angle $2\theta$ to the axis
meet it at points distant $a(1-\frac{1}{2} {\rm sec} \theta)$ from
the pole, i.e., at 
\begin{eqnarray*}
{\scriptscriptstyle{\frac{1}{2}}} a ({\rm sec} \theta -1) \equiv
x(\theta)\ , 
\end{eqnarray*}
from the paraxial focus of rays at small $\theta$.  We measure
$x(\theta)$ in the direction of the original parallel rays and here we
have taken the origin of $x$ at the paraxial focus.  While this choice is
convenient here, more generally we shall take the defocusing function
to be measured from the zero point of whatever coordinate system is in
use.
\begin{eqnarray*}
\lefteqn{U\arcmin(\theta) =g(\theta) = \left [x_f - {\scriptscriptstyle{\frac{1}{2}}} a({\rm sec}\theta
-1)\right ] \sin 2\theta =} \\ & & 
\left [2x_f \cos \theta -a (1-\cos \theta
)\right ]\sin \theta\ , \label{ex2} 
\end{eqnarray*}
\begin{eqnarray*}
U(\theta) = \int^\theta_0g(\theta)d\theta -x_f +x_c =
-x_f {\scriptscriptstyle{\frac{1}{2}}} (1+\cos 2 \theta) + \\  
+ a \left [
 {\scriptscriptstyle{\frac{1}{4}}} (1-\cos 2 \theta) -(1-\cos
 \theta)\right ] +x_c\ . \label{ex3}   & & 
\end{eqnarray*}
Thus $\tau = - {\scriptscriptstyle{\frac{1}{2}}} U\arcmin/U = $ 
\[
= - \frac{1}{2}
{x_f \sin 2 \theta -a (1-\cos\theta)\sin \theta \over
x_c-x_f \cos^2 \theta + a \ \left [\frac{1}{2}\sin^2 \theta -
        (1-\cos \theta)\right ]}\ . \label{ex4}\hfill{\rm e}
\]
The shapes of all possible correctors are now specified by
(\ref{copyeq1}) \& (\ref{copyeq2}) with $\rho(\theta)$ given by
(\ref{eq12}) and $\tau(\theta)$ by e(\ref{ex4}).  One merely plots
$R(\theta)$ against $X(\theta)$ to get the form of the corrector
mirror.
\begin{figure}
\epsfig{figure=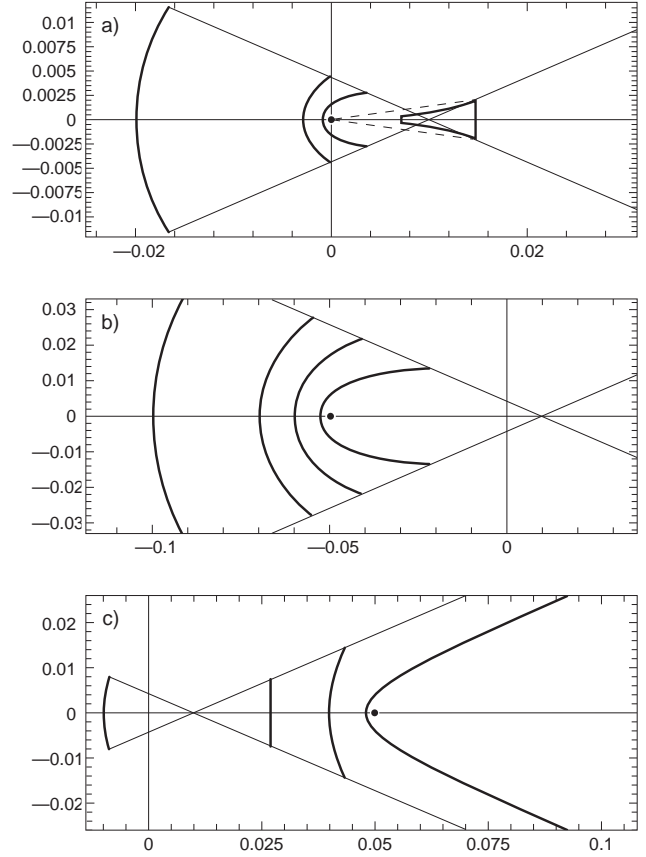}
\caption{Twelve possible single mirror correctors for a fast
sphere.  Only the rays from the edge of the primary are drawn.\ \ \ \
\ {\bf a)}  Four corrector mirrors that give a final focus at
the polar focus of the primary $(0,0)$.  The strange glancing
incidence cusped mirror collects 62\% of the light and obeys Abbe's
condition.  For this corrector alone, the reflected rays to the focus
are shown, dotted. See also Figure 3 for more detail.  
\eject 
{\bf b)}  Four corrector mirrors each giving a
final focus at $(-0.05,0)$.\eject  
{\bf c)}  Four corrector mirrors each
giving a final focus at $(0.05,0)$.} 
\label{fig2abc.eps} 
\hrule
\end{figure}
For a corrected focus at the original polar
focus at $(0,0)$, we take $x_f = 0$ and plot four possible
corrector mirrors as figure 2a.  Extreme rays at $2\theta = \pm 0.4$
radians are drawn.  The spherical aberration has them crossing at about
0.01 rather than at zero.  After hitting any one of these correctors
these (and all other rays) will be redirected to pass through the new
exact focus at $0$.

One of the mirrors is now very strangely shaped (like a trumpet), with
a pronounced cusp at the focus.  It reflects some rays by external
glancing incidence (actually 62\% of the light, see figure 3) but gets
in the way of those near the axis.  We shall say more of this later.

In figure 2b we draw four other correctors for the same spherical
primary but we have now moved the final focus to $-0.05$ to the left
of the original paraxial focus.  Again {\bf all} these mirrors
reflect the diverging rays to the new focus and all are possible
alternative correctors for spherical aberration.

Figure 2c is the same as the others, but now the final focus is to the
right of the uncorrected one, so all the mirrors reflect the 
rays back to make the corrected focus at $+0.05$.  

The diagrams illustrate the variety of corrector mirrors for just one
primary, a fast sphere.  There is a two parameter set of possible
correctors, however, most of them suffer from very bad coma when used
off axis.  Most of the rest of this paper is concerned with the
question of eliminating this coma by suitable choices of the shapes of
both primary and secondary mirrors.  We now show that the very strange
cusped mirror of figure 2a satisfies Abbe's condition and has no coma
although its peculiar shape gives it a very small field.

This unusual, but interesting, trumpet--shaped corrector is found by
taking $x_f = x_c = 0$ in e(\ref{ex4}).  Then
\[
g = U\arcmin =  
        -a (1-\cos \theta)\sin \theta\ ,  
\label{ex5} \hfill {\rm e}
\]
\[
U =  a(1-\cos \theta) \left [\frac{1}{2}(1+\cos \theta)-1\right ] \ ,
\label{ex6} \hfill {\rm e}
\]
\[
\tau = - \frac{1}{2} {U\arcmin \over U} = 
        {-\sin \theta \over 1-\cos \theta} = - \cot (\theta/2) = \tan
\left ({\theta \over 2} + {\pi \over 2}\right ). 
\label{ex7} \hfill {\rm e}
\]
So
\[
2\psi = \theta + \pi \ {\rm and}\ \phi = \pi - \theta \ .
\label{ex8} \hfill {\rm e}
\]
For such a corrector the rays are not reflected back but are given a
glancing reflection which allows them to continue to the focus, see
Figure 3.  In this case the focus coincides with the paraxial focus
of the uncorrected rays.  The shape of the corrector as given by (\ref{eq6})
and e(\ref{ex7}) is
\[
\rho = a (1+\cos \phi) = (1-\cos \Phi)\ ,\label{ex9} \hfill {\rm e}
\]
which is a cardiod with its cusp at the focus. We have defined the
acute angle $\Phi = \pi - \phi$ which for this system is equal to
$\theta$.  Thus after correction the rays halve their angle to the
axis and come into the focus 
at $\theta$ rather than $2\theta$.  Since the focus is $a/2$ from the
 mirror's centre of curvature, such rays are parallel to the normals to
the primary mirror at the points where they hit it.  Since the focus
is at $a/2$ beyond the centre of curvature, the extrapolated rays from 
focus meet the corresponding extrapolated incoming parallel rays on a
sphere $a/2$  behind the primary mirror.  Thus the sphere and the cardiod
corrector give an example of a perfect focus telescope obeying
Abbe's condition (see section 4).  The final image is therefore
free of coma.  We explore this system in another paper because it
leads to very small correctors for very fast spheres.  
\begin{figure}
\epsfig{figure=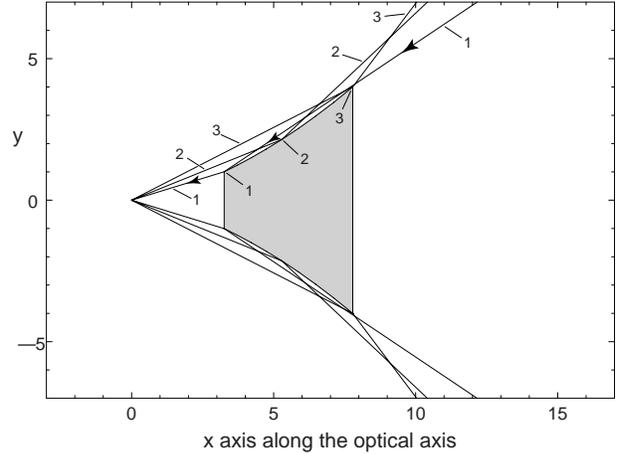}
\caption{This strange trumpet-shaped corrector for a spherical primary
removes both spherical aberration and coma. The highly convergent rays
are reflected glancingly from the outside of the corrector and on
reflection the rays have angles to the axis half as great as those
incident on the corrector. Only 62\% of the light is caught before the
corrector's surface gets in the way of the incident light but the
remaining 38\% can be redirected to form a separate focus. The sphere
and trumpet corrector form the $s={1\over 2},\  K=-2$ system of
section 4.  To allow the separate rays to be seen this figure is drawn
for a faster beam than figure 2a. }
\label{fig3.eps} 
\hrule
\end{figure}
\section{The Primary is determined by the defocusing function}
Clearly any primary mirror will have  a defocusing function, but in
this section we are interested in the inverse problem of determining
the shape of the primary mirror when defocusing function $x(\theta)$
is known and is due to just the primary.

Let $(X_p, R_p)$ be a point on the primary mirror as in figure 1.
From the geometry
\[
R_p = (X_p -x(\theta))\tan 2\theta\ , 
\label{eq16}
\]
and
\[
-dX_p/dR_p = \tan \theta = t \ .
\label{eq17}
\]
Equation 
(\ref{eq16}) may be rewritten
\[
\left [\left (1-t^2\right )/(2t)\right ]R_p = X_p - x\ .
\]
Differentiating with respect to $t$ and using (\ref{eq17}) 
 for $dX_p/dR_p$, 
\[
{1+t^2 \over 2t} \ {dR_p\over dt}\ -\ {1+t^2 \over 2t^2}R_p = - {dx
\over dt} \ ,
\]
and hence
\[
{d \over dt}\left ({R_p \over t}\right ) = - {2 \over 1+t^2}\ {dx
\over dt}\ .
\]
On integrating by parts and multiplying by $t$
\[
R_p = - {2t \over 1+t^2}\ \ x - 2t \left [\int^t_0 {2xt \over \left
(1+t^2\right )^2} dt-c_1 \right ] \ ,
\]
therefore 
\[
R_p = - x \sin 2\theta - 2 \tan \theta \left [\int^\theta_0 x\sin 2
\theta d\theta -c_1 \right ] \ . \label{eq18}
\]
This gives $R_p$ in terms of $\theta$, and $X_p$ is then determined
in terms of $\theta$ via (\ref{eq16}).   
Hence (\ref{eq18}) 
 and (\ref{eq16}) 
 give parametric equations for determining the primary mirror's shape.
Notice that this shape depends on the integration constant $c_1$.  We
may determine this by dividing (\ref{eq18}) 
by $\tan 2 \theta$ and then
letting $\theta\rightarrow 0$ the left hand side becomes $x_p - x(0)$
from (\ref{eq16}), with $x_p$ the position of the primary's pole.  The right
hand side becomes $-x(0) + c_1$ so $c_1 = x_p$ and the only parameter
is thus the position of the primary.

We found that $U(\theta)$ was preferable to $x(\theta)$ at least
for the specification of the corrector.  It is therefore of interest
to re-express the equations for the primary in terms of $U(\theta)$
rather than $x(\theta)$.  From (\ref{eq10})
\[
x(\theta) = x_f - U\arcmin (\theta)/\sin 2\theta \ ;
\]
using this and expression (\ref{eq10}) for $U$ in (\ref{eq18}) we
find
\[
R_p = +U\arcmin + 2\tan \theta
(U+x_p-x_c)\ . \hfill\label{eq19}\footnote{1}
\]
\footnotetext{when $x_f < x_c$
(\ref{eq19}) must have $x_p-x_c $ replaced by $x_p + x_c -2x_f$}
Thus, if the function $U(\theta)$ is given, the mirror separation
$x_p-x_c$ will give us $R_p(\theta)$, and $X_p(\theta)$ is given by
(\ref{eq16}) 
\[
X_p = R_p \cot 2\theta - U\arcmin/\sin 2 \theta +x_f\ . \label{eq20} 
\]
A number of corrector mirrors for a fast sphere are illustrated in
figure 2 a, b, c.  The rays are drawn for the strangest of these in
figure 3.  Just as for the secondary, the shape of the primary is
determined by $U(\theta)$ but it now needs also the mirror separation.
The parameter $x_f$ is needed only if we want to locate the two--mirror
system within our arbitrary coordinate system.  Thus $U(\theta)$
together with the separation $x_p - x_c$ specifies the 2 mirrors.
Notice that we can, if we so wish, choose a coordinate system zeroed
at final focus, in which case $x_f = 0$ and that parameter vanishes
from the system.

Glancing incidence primaries can be included in the same analysis by
letting $\theta$ take the values between $\pi/4$ and $\pi/2$. 
\section{The Basic Theorem on Two Mirror Systems}
We have shown that the uniting function $U(\theta)$ together with the
mirror separation determines the shapes of both mirrors.  We now show
how the equal magnification or no coma condition of Abbe can be
used to determine $U(\theta)$.  Abbe's condition (Abbe 1873)
is that rays approaching the final focus when extrapolated back to
meet their corresponding incoming parallel rays must intersect them on
a sphere centred at the focus.  In our notation this means
\[
R_p = b\sin \phi \label{eq21}
\]
where $b$ is the radius of the  Abbe sphere which is the effective
focal length of the optical system.
\subsection*{Theorem}
In any two mirror telescope with no spherical aberration and no coma,
the angle of the rays to the axis after reflection in the primary,
$2\theta$, is related to the angle $\phi$ at which those rays enter
the final focus by
\[
t={1\over s}\left [T/(1+T^2)\right ]\left [1-K \vert 1-T^2/\eta
\vert^{-\eta}\right ]\ ,
\]
where $t=\tan\theta,\ T=\tan \phi/2,\ \eta = s/(1-s)$ and $s$ and $K$
are constants which give the mirror separation and position of final
focus as fraction of the effective focal length, $b$.  Furthermore,
the shape of the primary mirror $R_p(X_p)$ is given parametrically
in terms of $T$ and $t(T)$ by
\[
R_p     =       2bT/(1+T^2)\ ,
\]
\[
X_p     =  b \left \{ s - (1+T^2)^{-1} + (t/T) \left
[s-T^2/(1+T^2)\right ]\right \} \ ,
\]
where the origin has been chosen at the final focus, $(x_f = 0)$.
Finally the corrector or secondary mirror $R(X)$ is given
parametrically by
\[
R = \rho \ 2T/(1+T^2)\ ,
\]
\[
X =     -\rho (1-T^2)/(1+T^2)\ ,
\]
where
\[
\rho =          bK\vert 1-T^2/\eta
        \vert^{-\eta}/(1-tT)\ .
\]
\subsection*{Proof}
Rewriting (\ref{eq19}) in terms of $U$ by using (\ref{eq5}) and
(\ref{eq11}) we find, on simplification
\[
R_p = -2U\left [\tan \left (\theta + \phi/2 \right ) - \tan\theta
\right ] + 2 \left (x_p - x_c\right ) \tan \theta\ . \label{eq22}
\]
(\ref{eq21}) 
 and (\ref{eq22}) 
 allow us to express $U$ in terms of $\theta, \phi, b$
and the dimensionless separation of the mirror $s$
\[
s = (x_p - x_c)/b\ . \label{eq23}
\]
Thus
\[
U=b \left (s\tan \theta - {1 \over 2} \sin \phi \right ) / \left [
\tan (\theta + \phi/2) - \tan \theta \right ]\ . \label{eq24}
\]
Now write $\tan \phi/2 = T$ and $\tan \theta = t$ as previously and so
obtain 
\begin{eqnarray*}
\tan (\theta + \phi/2) - \tan \theta = \left [t+T - (1-tT)t\right
]/(1-tT)   = \\
= T (1+t^2)/(1-tT).
\end{eqnarray*}
Using this in (\ref{eq24}) we find with (\ref{eq21})
\[
U=b\left [stT^{-1}-(1+T^2)^{-1}\right ](1-tT)/(1+t^2)\ .\label{eq25}
\]
Differentiating ${\rm ln}U$ with respect to $T$ we find

\begin{eqnarray}
U^{-1} dU/dT = {-stT^{-2}+2T(1+T^2)^{-2} \over stT^{-1} -
(1+T^2)^{-1}}{t \over 1-tT} + \nonumber \\ 
+ {dt \over dT} \left \{ {sT^{-1} \over
stT^{-1} - (1+T^2)^{-1}}- {T \over 1-tT}-{2t \over 1+t^2}
\right \}\ . \label{eq26}
\end{eqnarray}
But, by (\ref{eq11}) and (\ref{eq5})
\begin{eqnarray}
U^{-1} dU/dT \cdot dT/dt \cdot dt/d\theta = U^{-1} dU/d\theta =
\nonumber \\
= -2\tan (\theta + \phi/2)  
 = -2 (t+T)/(1-tT)\ . \label{eq27}
\end{eqnarray}
Dividing by $(dT/dt)(1+t^2)\equiv dT/dt \cdot dt/d\theta$ we find
\[
U^{-1} dU/dT = -2        (dt/dT)(t+T)/\left [(1+t^2)(1-tT)\right ]
\]
with which we replace the left hand side of (\ref{eq26}) 
 to give a
differential equation for $t(T)$.  In this the last two terms in the
coefficient of $dt/dT$ on the right combine with those from the left
to give $+T/(1-tT)$.  Multiplying by $T\left
[stT^{-1}-(1+T^2)^{-1}\right ](1-tT)$ a miracle occurs in that the
resulting equation is linear in $t$, viz 
\[
{dt\over dT}\left(s-{T^2\over1+T^2}\right)-{t\over T}
\left[s-{T^2 (1-T^2)\over(1+T^2)^2}\right]={-2T^2\over
(1+T^2)^2}.\label{eq28}
\]
Such equations have integrating factors, $I$, and (see Appendix),
\[
I=T^{-1}(1+T^2)\vert 1-T^2/\eta \vert^\eta,\ 
\]
where
\[ \eta =
s/(1-s),\ s=\eta/(\eta +1)\ .  \label{eq29}
\]
(\ref{eq28}) 
 becomes,
\[
{d(It)\over dT} = - 2 \iota s^{-1} T \vert 1-T^2/\eta \vert^{\eta
-1}\ , 
\]
where $\iota = \pm 1$ according as $1-T^2/\eta$ is positive or
negative.  On integration, we find for $\eta \neq 1$ 
\[
t={1\over s}\left [T/(1+T^2)\right ] \left [ 1-K\vert 1-T^2/\eta
\vert^{- \eta} \right ]\ , \label{eq30} 
\]
where $K/s$ is the constant of integration.  

Q.E.D.

Using this expression for $t$  in the first factor of (\ref{eq25})
\[
U=-bK {(1-tT) \vert 1-T^2/\eta \vert^{-\eta}\over (1+t^2)(1+T^2)}\ .
\label{eq31}
\]

Now by (\ref{eq10}) $U \longrightarrow -(x_f -x_c)$ as $\theta
\longrightarrow 0$ and assuming $T \longrightarrow 0$

\[
bK = x_f - x_\iota.  \ {\rm Hence}\ K=(x_f - x_c)/b\ . \label{eq32}
\]
Thus $K$ determines the amount by which the focus lies downstream of
the corrector in terms of the effective focal length $b$.
(\ref{eq30}) 
 in the
light of (\ref{eq29})
depends on only two parameters $K$ and $s$.  These are
the distance from the corrector mirror to the final focus and the
mirror separation both measured in terms of the effective focal length
$b$.  

The drawing and the limit on which the interpretation is based have
assumed that when $\theta \longrightarrow 0,\ T \longrightarrow 0$.
This is correct when the corrector mirror reflects the rays towards
the primary mirror but is incorrect for the glancing reflection
corrector for the sphere considered in the example of section 2.  For
such systems $\phi \longrightarrow \pi - \theta$ so $T = \tan \phi/2
\longrightarrow \infty$ as $\theta \longrightarrow 0$.  Looking at
(\ref{eq30}) 
we find that case is given by taking $\eta =1$ $(s={1\over 2})$  and $K = - 2$ so
that as before
\[
t=-2T/(1-T^2).
\]
Returning to the normal case but with $T$ and $t$ both small
\[
t\simeq \left ({1+r \over \eta}\right )\left (1-K \right )T\ {\rm
so}\ \theta = \left ({1-\eta \over \eta}\right ) \left (1-K\right )
\phi/2 \ . 
\]
we therefore see that $\eta$ small gives $ \theta \gg \phi$ while
$\eta$ large gives $\theta \sim (1-K){1\over 2}\phi$.

We notice that $\theta$ is also small when $T^2 \simeq \eta \left [
1-K^{1/\eta} \right ]$.  When $K=1$ and $T$ is small $t= \left
\{(1-\eta)/\eta \right ]T^3$ so $\phi = -2 \left[\eta
\theta/(1+\eta)\right ]^{1/3}$.

\subsection{The Forms of the Mirrors}
With $U$ given by (\ref{eq31})
, $U\arcmin$ may be found from (\ref{eq27})
\[
U\arcmin = 2b K {(t+T) \vert 1-T^2/\eta \vert^{-\eta} \over 
(1+t^2) (1+T^2)}\ , \label{eq33?}
\]
hence from (\ref{eq12})
\[
{\rho \over b} = K {\vert 1-T^2/\eta \vert^{-\eta} \over
1-tT} \ ,
\]

with $\rho$ known as a function of $T$ (\ref{eq14}) \& (\ref{eq14a}) 
determine $R$ and $X$
parametrically as functions of $T$.  $x_f$ merely gives a zero--point
shift which we may take zero so the shape of the corrector is known.

For the primary (\ref{eq21}) 
gives $R_p = 2bT/(1+T^2)$.  Using this and (\ref{eq24}) 
in (\ref{eq20})
we find
\[
X_p - x_f = b \left [s- {1 \over 1+T^2} + {t \over T} \left
(s - {T^2 \over 1+T^2}\right )\right ]\ ;
\]
this gives a parametric equation in terms of $T$ for the primary
mirror.  This completes the proof of the second half of the theorem
i.e., that which gives the mirror shapes.
\section{Singularities Asymptotes and Catastrophes in the
  Mirrors}
After equation (\ref{copyeq2})
 we commented that because the equations for the
corrector mirror's shape are parametric, cusps and singularities can
be expected. Indeed the general theory of the shapes obtained when
the parameters of parametric equations are eliminated is called
Catastrophe Theory and not without reason. Poincar\'{e} was the first
to develop it and he applied it both in orbit theory and to the
theory of rotating liquid masses. More recently the theory was
much developed and elaborated by Thom, Zeeman and Arnold.
Both mirrors of the coma and spherical aberration free optical
systems are defined by parametric equations so we can expect cusps
to occur at special points of each of our mirrors for some $s,\ K$ values.
 Also it is quite possible for a mathematically defined mirror to reach
out to infinity and to come back from the other direction as such
asymptotes are common in mathematics. In practice only a small
part of a mathematically designed mirror is made and much has
to be left out anyway to let the light see the pieces of mirror
that are to be used.

     Since the special points where cusps or asymptotes occur
in the mirror shapes can be determined from a study of the equations
that define the mirrors, we now step aside from the main line
of this paper to make such a study. Of course it is possible just
to draw the shapes of the mirrors from the parametric equations
and so discover the singular points merely by observation. Those
less interested in mathematics may be well advised to go straight to
figure 4 which describes the special points encountered on the
mirrors for each value of $s$ and $K$. The theory that precedes that
figure allows the reader to understand why the boundaries between
mirrors with different types of cusps or asymptotes occur where
they do in the $s,\ K$ parameter space.

$R_p = 2bT/(1+T^2)$ maximises at $b$ when $T=1$, that is $\phi =
\pi/2$.  At that $\phi$ the rays approach the final focus at right
angles to the axis.  Now from (25) $dX_p/dT = -t dR_p/dT$ so $dX_p/dT$
is also zero at $T=1$.  Both $R_p(T)$ and $X_p(T)$ turn back at the
same point which makes a cusp in the primary mirror.  These cusps
point in the direction of increasing $x$ when $t(1)<0$ and decreasing
$x$ when $t(1)>0$ where $t(1)= {1\over 2s}\left [ 1-K\vert 2-{1\over
s}\vert^{s/(s-1)}\right ]$.  Beyond this cusp the rays enter the final
focus backwards $(\phi > \pi/2)\ T>1$ and we refer to the primary as
having its second sheet, see Figure 5a.  Since $R_p$ remains in the
range $0\leq R_p \leq b$ for all $T$ its radius is always finite,
however this does not mean that the mirror is always finite 
because it can reach out to infinity in directions parallel to the
axis.  We may rewrite (37)
$$X_p = bs -b \left [1-(st/T)\ (1-T^2/\eta)\right ]/(1+T^2)\ .$$
Using (35) for $(st/T)$ we see that $\vert X_p\vert \longrightarrow
\infty$ when and only when
$(1-T^2/\eta)^{1-\eta}/(1+T^2)^2\longrightarrow \infty$.  This occurs
for $\eta >1$ when $T^2=\eta$ and for $\eta <-1$ when
$T\longrightarrow \infty$.  The latter case gives a singular backward
$(x\longrightarrow -\infty)$ spike on axis for the primary's second
sheet when $s>1$.  The former case also occurs on the second sheet
since $T^2=\eta >1$.  The mirror then asymptotes to the cylinder $R_p =
2b\eta^{1/2}/(1+\eta)$.  For $K>0$ the mirror approaches this cylinder
from the outside ($R_p$ decreasing) as $T$ increases towards $\eta$
and $X_p \longrightarrow - \infty$.  The mirror then reappears at
large positive $X_p$ and decreasing $X_p$ with $R_p$ still decreasing.
For $K<0$ the $X_p$ behaviour is reversed.

As $T\longrightarrow \infty$ the second sheet of the primary
approaches the axis.  From (25) $dX_p/dR_p = -t$ so there will only be
a regular pole to the mirror's second sheet if $t\longrightarrow 0$ as
$T\longrightarrow \infty$.  We already saw that $\eta<-1,\ s>1$ the
mirror asymptotes to a spike on axis but from (35) $t(\infty)$ is
still not zero for $-1<\eta\leq - 1/2$.  In this range there is a cusp
on axis.  For $\eta = -1/2 $ it is an open cusp with $dR_p/dX_p$
approaching the finite gradient $\sqrt{2} s/K$, but for the rest of
that range it is a sharp cusp with $dR_p/dX_p = 0$ there.  

The corrector mirrors also have asymptotes unless $K/s > 0$ with
$\eta$ negative (i.e., $s$ outside the range 0 to 1).  Since $R/X =
\tan \phi$ these asymptotes are caused as $\rho \longrightarrow
\infty$ and make cones with apices at the final focus.  From (40) it
seems as though there would be infinities in $\rho$ when $T^2=\eta >0$, 
but from (35) we see that $t$ has a compensating infinity there, so
the only infinities in $\rho$ are when $1-tT$ is zero.  Using (35) for
$t$ and the relation between $s$ and $\eta$ these occur when
$$(K/s)T^2 = -\vert 1-T^2/\eta\vert^\eta (1-T^2/\eta)\ . $$
As stated earlier such asymptotes occur unless $K/s$ is positive and
$\eta $ negative in which case the two sides have opposite signs.  All
other cases have asymptotes.

The occurrence of cusps in the corrector mirror is best considered in
polar coordinates $\rho, \phi$ except that we use $T=\tan \phi/2$
rather than $\phi$.  Combining equations (40) and (35) we have the
following expression for $\rho(T)$.
$$\rho/b =K(1+T^2)/\left [(1-T^2/\eta)\vert 1-T^2/\eta\vert^\eta +
KT^2/s \right ]\ .$$
The zeros on the denominator give the asymptotes that we have just
discussed.  The gradient change due to the modulus sign is always
eliminated by the zero of its factor $(1-T^2/\eta)$ so does not lead
to a cusp.  Thus the only cusps occur at focus, $\rho=0$ with
$T\longrightarrow \infty$ and when $\eta >0$, i.e., $s$ in the range
$(0,1)$.  For $\eta <0,\ \rho \longrightarrow bs$ as $T\longrightarrow
\infty$ and the secondary has a smooth regular pole just as it does
when $T\longrightarrow 0$.  

The results of this section are summarised in Figure 4.
\begin{figure}
\epsfig{figure=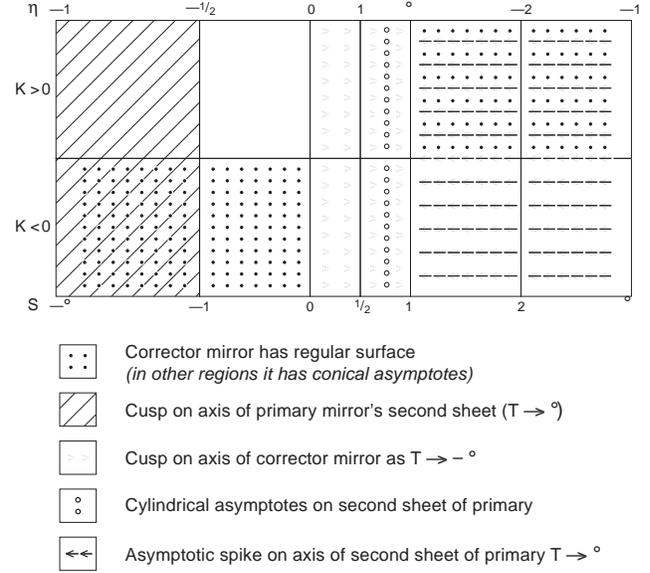}
\caption{The diagram illustrates by different shadings, which
  sometimes overlap, the parts of the $s,\ K$ plane of optical designs in
  which the mirrors have different types of singular behaviour. For
  example a design with $s <-1$ and $K < 0$ will have both a corrector
  mirror without singularities and a primary with a cusp on the axis
  of its second sheet. This is why such designs lie in regions in
  which the diagonal shading overlaps the dots in squares pattern. The
  values of $\eta = s/(1-s)$ are given at the top for convenience.}
\label{fig4.eps} 
\hrule
\end{figure}

\section{Use of the Formulae}
Mathematics can be blind and misleading.  We have not required that
the light can get to the primary without hitting the secondary, nor
that the light can actually reach the focus.  Some of the foci are
virtual, sometimes it is only the backward extrapolation of the ray
that is reflected from the primary that would have hit the secondary
and been reflected to the focus.  Thus, while we have all useful
spherical--aberration--and--coma--free two mirror systems, we have to cut
away unused pieces of the mirrors to get them and along with useful
designs there are many useless ones.  There is no substitute for
drawing the designs in detail, so that one may ensure that the system
can be baffled against stray light.  This is done in the companion
paper by Willstrop and Lynden-Bell (2002).

In figures 1--6 we assumed that the light came in from the left
initially, hit the primary to the right of the diagram, and returned
to hit the secondary.  In applying the formulae one finds that the two
parameter set of designs is best studied in the $s-K$ plane and that
when $s$ is negative, the light should be assumed to come in
from the right instead.  However, we like to maintain the convention
that the light comes in from the left so we have left-right reversed
figure \nolinebreak7.

Most of the good designs are perfect focus variants of well known ones
listed in Table 2 with the values of $K$ and $s$ that give them.
\begin{table}
\caption{$s$ and $K$ values for some designs}
\begin{tabular}{lll}
& $s$ & $K$\\
Ritchey--Chr\'{e}tien (1922)& $0.274$  & $+0.335 $       \\
Schwarzschild       (1905)      & $1.25$ & $+0.50 $      \\
Couder              \parbox{3cm}{(see Willstrop '83,84)}        
                        & $2.00$ & $+0.385$             \\
$X-$Ray Telescopes &$0.0008$     & $-1.05$ \\
Sphere corrected by ``trumpet''
                  & $0.5$         & $-2.0$\\
Bowen Spectrograph Camera
                 & $2.0$           & $+4.236$\\
Solar Furnace    & $-2.0$    & $+0.1$    
\end{tabular}
\hrule
\end{table}
\begin{figure}
\epsfig{figure=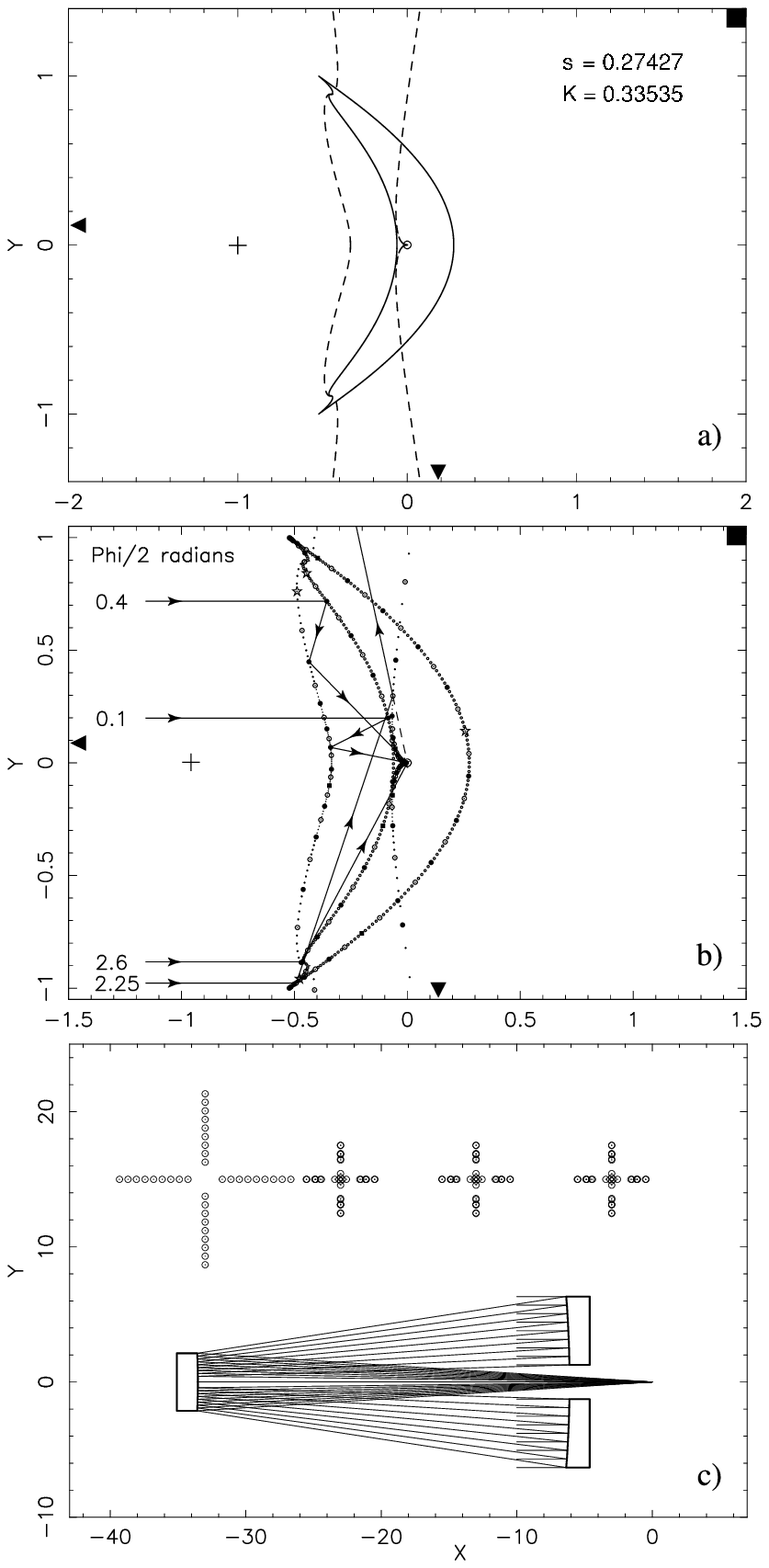}
\caption{\eject
{\bf a)}  Mirror profiles for the `perfect--focus' version of a
Ritchey--Chr\'{e}tien system $s=0.274,\ K=0.335$.  The two sheets of a
primary are drawn with a continuous line with second sheet on the
right.  The corrector is drawn dashed and has an asymptote.  There is
also a cusp on the axis of the corrector mirror's second sheet
$T\longrightarrow \infty$.  \eject
{\bf b)}  As 5a with some rays drawn.  That with $\phi/2 = 2.25$
gives only a virtual focus.  \eject
{\bf c)} Parts of the mirrors used in the Anglo Australian
Telescope.  }
\label{fig5abc.eps} 
\hrule
\end{figure}
Figure 5 gives the $s = 0.274,\ K= 0.335$ perfect focus version
Ritchey--Chr\'{e}tien telescope design.  In 5a we have drawn  just the
mirrors for all values of $T$.  The primary mirror has the full line.
Its profile makes a strange wiggle just prior to the edge cusp.  The
second sheet is regular and remains smooth on axis.  The corrector is
drawn dashed; proceeding upwards on its (left-hand) first sheet it
also has a wiggle but then asymptotes upper left to infinity
reappearing lower right on its second sheet.  This second sheet meets
the axis in a cusp (in accordance with figure 4).  In figure 5b the
paths of some representative rays are illustrated.  Clearly large
parts of the secondary must be cut away to let the light hit the
primary.  The ray at $\phi/2 = 2.25$ radians is reflected away from
focus by the secondary and so gives only a virtual focus.  The focus
is marked by a larger circle.  In figures 5b, 6a and 7 the primary is
represented by a chain of small circles with every fifth one larger
and every tenth filled in.  The secondary is denoted by a chain of
dots with every fifth a circle and every tenth filled.  Virtual rays
are dashed.  

In the lower half of Figure 5c just the useful parts of the mirrors
are shown along with a number of rays starting at the plane $X= -0.1$,
initially parallel to the axis, falling on the primary and secondary
mirrors in turn, and ending at the focus at $X = 0.0$.  In the upper
half of Figure 5c are four spot diagrams.  At the left, we show the
distribution of rays on the primary mirror.  The other three spot
diagrams are greatly enlarged, and show the computed distribution of
rays at the focus.  In this case the image is on axis, and all three
of these spot diagrams are identical.  Their size is entirely due to
the limited accuracy of our computer, which uses double precision
codes.  The total spread of each image is under $3 \times 10^{-14}$
radians, or $6 \times 10^{-9}$ arc-seconds.
\begin{figure}
\epsfig{figure=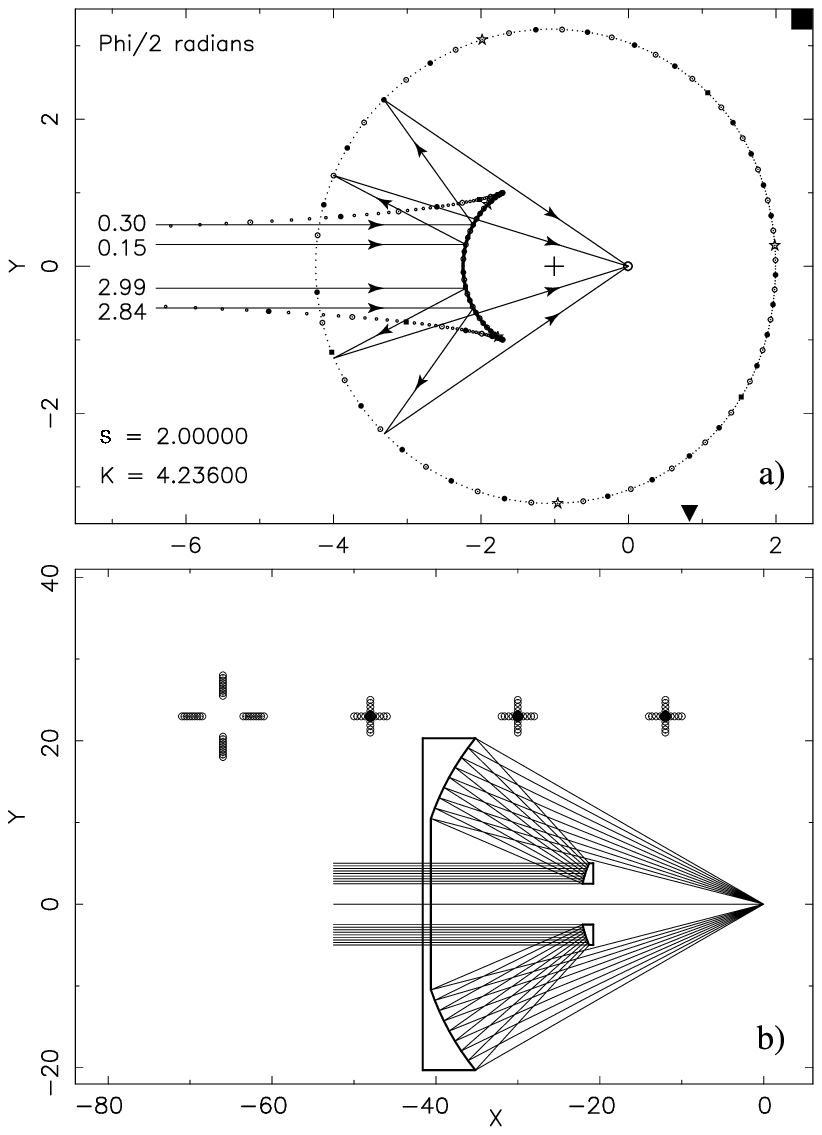}
\caption{{\bf a)}  The $s=2.0 \ K=4.236$ Perfect--Focus System gives
the Bowen spectrograph camera.  Notice that the primary has an asymptotic spike
on axis of its second sheet and the corrector is regular everywhere.  \eject
{\bf b)}  Ray paths and mirror parts used in the camera.}
\label{fig6ab.eps} 
\hrule
\end{figure}
\begin{figure}
\epsfig{figure=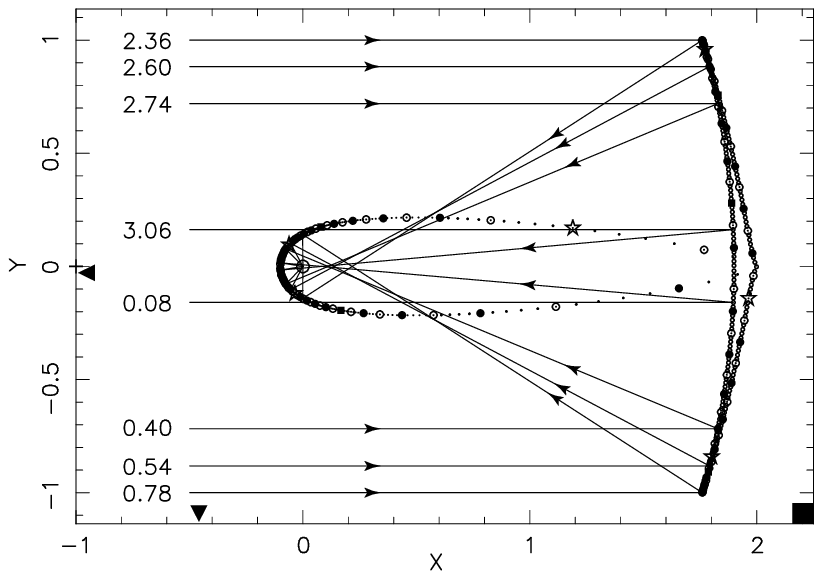}
\caption{The $s=2.0,\ K=-0.1$ Perfect--Focus System gives a focus at
which the rays enter over $2\pi$ steradians.  Such systems are
suitable for solar furnaces.}
\label{fig7.eps} 
\hrule
\end{figure}
Figures 6 and 7 are included both to illustrate the catastrophe types
of figure 4 and to demonstrate the great variety of perfect focus
systems encapsulated in our formulae.  

Figure 6 gives a perfect focus
version of the Bowen spectrograph camera while figure 7 gives a design
in which rays approach the final focus over a full hemisphere.  Such
coma--free designs are more suitable for solar furnaces or light-house
beams than they are for telescopes or spectrographs.

Superficially similar designs are illustrated in Mertz (1996) who says
the relevant figure originated in the work of Sebastian von Hoerner.  However,
on closer inspection, these designs are correctors for a spherical
primary and thus cannot obey Abbe's no coma condition exactly.  In
spite of this these corrector designs for a sphere may be more
practical than those given here with an aspheric primary because the
corrector assembly can be moved to follow the sun for a significant
time even if the spherical primary is fixed.  For an aspheric primary
such a movement would destroy the spherical aberration correction.
Mertz gives there a number of other designs for solar furnaces.

There are also designs for systems which have either primary or
secondary of very low power.  One of these gives an all reflecting
Schmidt which replaces Schmidt's corrector plate by a reflector which
gives the returning rays just the right corrections for the sphere.
Only by turning a normal Schmidt upside-down and punching a large hole
in the spherical mirror could any light enter this system and its
efficiency would then be very low. It may be that other `almost flat'
mirrors could be better replaced by suitably figured glass as in the
real Schmidt.   However, there are variants of the
Strand astrometric telescope at Flagstaff that slightly figure the flat
secondary and greatly improve the field.

In the accompanying paper (Willstrop \& Lynden-Bell, 2002), we
describe how to use the parametric equations of the mirrors to trace
off axis rays.  We also discuss the designs found over the whole $s-K$
plane from which the above are a selection.

\section*{Acknowledgements}
It is a pleasure to thank Dr R.V. Willstrop for his unflagging
enthusiasm and support.  Roger Angel and Richard Hills helped to
introduce me to the literature.  Prof. R. Stobie, Director, SAAO
supported a visit that generated this work.

\appendix
\section[]{Evaluation of the Integrating Factor}
\begin{eqnarray}
d{\rm ln}I/dT^2 = {s(1+T^2)^2 -T^2(1-T^2) \over
                        2T^2(1+T^2)\left (s+T^2(s-1)\right )}\nonumber
= {- 1/2 \over T^2} + \\
+ {1 \over 1+T^2} - 
                {s \over s+T^2(s-1)} \ ,
\end{eqnarray}
hence
\[
I=T^{-1}(1+T^2)\vert 1-T^2/\eta \vert^\eta \ ,
\]
where
\[
\eta = {s \over 1-s},\ s = {\eta \over \eta +1} \ .
\]
\bsp

\label{lastpage}

\end{document}